\documentclass[aps,preprint,showpacs,amsmath,amssymb,prd,nofootinbib]{revtex4-1}
\usepackage{amssymb,graphics,graphicx,epstopdf,color,slashed}

\begin{document}

\title
{\Large \bf Vacuum stability, neutrinos, and dark matter}

\author{Chian-Shu~Chen\footnote{chianshu@phys.sinica.edu.tw} and Yong Tang\footnote{ytang@phys.cts.nthu.edu.tw}}
\affiliation{Physics Division, National Center for Theoretical Sciences, Hsinchu, Taiwan 300}

\date{\today}

\begin{abstract}
Motivated by the discovery hint of the Standard Model (SM) Higgs mass around 125~GeV at the LHC,  
we study the vacuum stability and perturbativity bounds on Higgs scalar of the SM extensions including neutrinos 
and dark matter (DM). Guided by the SM gauge symmetry and the minimal changes in the SM Higgs potential we 
consider two extensions of neutrino sector (Type-I and Type-III seesaw mechanisms) and DM sector 
(a real scalar singlet (darkon) and minimal dark matter (MDM)) respectively. The darkon contributes positively 
to the $\beta$ function of the Higgs quartic coupling $\lambda$ and can stabilize the SM vacuum up to high scale. 
Similar to the top quark in the SM we find the cause of instability is sensitive to the size of new Yukawa 
couplings between heavy neutrinos and Higgs boson, namely, the scale of seesaw mechanism. MDM and Type-III 
seesaw fermion triplet, two nontrivial representations of $SU(2)_{L}$ group, will bring the additional positive contributions 
to the gauge coupling $g_{2}$ renormalization group (RG) evolution and would also help to stabilize the electroweak 
vacuum up to high scale.

%\noindent
\end{abstract}

\maketitle

\section{Introduction}
The mechanism of electroweak symmetry breaking has been searched for decades in past accelerators and on-going 
experiments at the Large Hadron Collider (LHC). In the standard model(SM), a scalar $SU(2)_L$ doublet is responsible 
for the symmetry breaking. The intrinsic problems associated the SM has intrigued the expectation of new physics at TeV 
scale. However, no evidence of new physics has been found so far at the LHC with $\sqrt{s}=7$ TeV. At the same time, 
both the ATLAS and CMS collaborations\cite{ATLAS2012,CMS2012} have reported on the search of SM Higgs with 
$\sim$5 fb${}^{-1}$ data and the results show a first hint of SM Higgs with $m_h=125\pm1$ GeV. This motivates us to 
study its implications in vacuum stability and perturbativity bounds within the extension of the SM including neutrino 
physics and dark matter (DM). 
%give a detailed survey of standard model taking other non-collider experimental evidences into account.

In the SM, the stability of the vacuum is tightly related with the mass of physical Higgs since the quartic self-coupling 
$\lambda$ is connected with $m_{h}$, $m_{h}=\sqrt{2\lambda}v$($v=246$ GeV). If $m_{h}$ is too small, the radiative 
corrections, mainly from the top quark's contribution, can drive $\lambda$ negative, induce a false and deep minimum 
at large field values and destabilize the electroweak vacuum. In terms of no clue to new physics beyond the SM at the 
LHC, the analysis of the stability for $m_{h}=125$ GeV can give useful hints on the structure of ultraviolet 
scale where the new physics may come in. 

On the other hand, non-collider experimental results confront the SM with two major puzzles: neutrino masses and dark 
matter. The phenomenon of neutrino oscillations show that at least two neutrinos have nonzero but small masses located 
around sub-eV scale. Evidences from astrophysics and cosmology have pointed out that the ordinary baryonic matter is not 
the dominant form of material in the Universe. Rather, about $23\%$ of energy density of the Universe is non-luminous and 
non-absorbing matter, called dark matter (DM). 

Enclosing dark matter and the massiveness of neutrino into SM may have effects on the Higgs sector for the analysis of 
stability of vacuum. Although the exact nature of dark matter and neutrino mass is still unknown and their interactions with 
SM particle vary for different models, there exist several guidelines for our purpose to analyze the vacuum stability. Based on SM gauge structure and changing the SM Higgs potential in a controllable way, the model should introduce new particles as 
less as possible to make the analysis of stability possible and necessary.

This paper is organized as follows. In section II we introduce two scenarios of the neutrino and the DM respectively, and 
explain why the our studied frameworks change minimally the scalar potential of the SM. Then we analyze the vacuum stability and perturbativity bounds in these frameworks. In section IV we give a summary and conclude our results.    

\section{The frameworks of neutrino and dark matter}
%We 

\subsection{The Type-I and Type-III seesaw mechanisms} 
Within the context of the SM, no Dirac mass term of neutrino can be written due to the absence of right-handed neutrino 
fields. If the neutrinos are Majorana fermions, that is, the lepton number is no longer a conserved quantity, one 
can write a dimension-five operator which is relevant to the neutrino masses~\cite{Weinberg:1979sa}  
\begin{eqnarray}\label{numass}
{\cal L}_{eff} = \frac{y_{ab}}{M}(\emph{l}_{L\alpha}H)(\emph{l}_{L\beta}H) + {\rm h.c.},   
\end{eqnarray}
where $l_{L(\alpha,\beta)}$ are the $SU(2)_{L}$ leptonic doublets with the flavor indices $\alpha, \beta = e, \mu, \tau$, and $H$ is the Higgs field. This operator violates lepton number by two units ($\Delta L = 2$), and hence M corresponds 
to the lepton number breaking scale. After electroweak symmetry breaking, the Higgs field develops vacuum expectation 
value (VEV), $\langle H\rangle = v/\sqrt{2}$ with $v = 246$ GeV, and the neutrino masses $m_{\nu} \sim yv^2/M$ 
are generated. In order to obtain the neutrino masses at sub-eV scale one either assume $M \gg v$ with a sizable 
couplings $y_{ab}$ or assume $M$ is achievable in the collider experiments with highly suppressed couplings $y_{ab}$. 

One of the most popular approaches to generate the neutrino masses corresponding to Eq.~(\ref{numass}) at 
tree level is the so-called "seesaw mechanism". In this kind of models the neutrino masses are suppressed by a large 
factor due to the masses of heavy sectors. The heavy fields are either fermions or scalars, therefore, three variations 
of seesaw mechanisms are introduced. In the so-called Type-I~\cite{seesaw} and type-III~\cite{Foot:1988aq} seesaw 
mechanism models, the leptonic $SU(2)_{L}$ singlet and triplet fermions are introduced respectively. Instead of introducing 
new fermions to the SM, the Type-II seesaw model~\cite{seesaw2} uses a $SU(2)_{L}$ triplet scalar carrying a 
hypercharge $Y=2$ to give neutrino masses through the new Yukawa interactions between the new triplet scalar 
and the SM leptonic doublet fields. This triplet scalar brings additional six parameters to the scalar potential which 
makes the analysis of vacuum stability unclear and complicated. So for the purpose of this paper, we concentrate on 
Type-I and Type-III seesaw mechanisms and discuss their effects on the vacuum stability bounds. 

\emph{Type-I seesaw mechanism}: One adds $N$ $SU(2)_{L}$ singlet fermions to the SM particle content, usually these 
particles are treated as the right-handed neutrinos $\nu_{R_{i}} (i=1,\cdots, N)$. Since $\nu_{R}$ fields do not carry any SM 
quantum numbers, one can write the relevant lagrangian in the neutrino sector as 
\begin{eqnarray}
{\cal L_{\nu}} = Y_{\nu_{\alpha i}}\bar{l}_{\alpha}\tilde{H}\nu_{R_{i}} + \frac{1}{2}M_{R_{i}}\overline{(\nu_{R})^c}_{i}\nu_{R_{i}} + {\rm h.c.}, 
\end{eqnarray} 
where $\tilde{H} = i\sigma_{2}H^*$, $Y_{\nu_{\alpha i}}$ is the Dirac Yukawa couplings, $c$ is the charge conjugation, and 
$M_{R_{i}}$ are the Majorana masses in the diagonal basis of right-handed neutrinos. After block diagonalizing the mass 
matrix of left and right handed neutrinos, one obtained the standard seesaw formula for the effective light neutrino 
mass matrix, 
\begin{eqnarray}
m_{\nu_{\alpha\beta}} = - \sum_{i = 1}^{N}\frac{M_{D_{\alpha i}}M^{T}_{D_{i\beta}}}{M_{R_{i}}}, 
\end{eqnarray}
here $M_{D}$ is the Dirac mass matrix of neutrinos. Radiative corrections from $Y_{\nu}$ can contribute the Higgs 
effective potential and have an effect on the vacuum stability.

\emph{Type-III seesaw mechanism}: Instead of the exchange of heavy fermion singlet fields in the Type-I seesaw model, 
Type-III seesaw mechanism is mediated by heavy fermions $\Sigma_{R}$ which have zero hypercharge and are transformed 
as a triplet in the adjoint representation under the $SU(2)_{L}$ gauge group. The triplet can be written in the tensor form,  
\begin{eqnarray}
\Sigma_{R} = \left(\begin{array}{cc}\Sigma^0_{R}/\sqrt{2} & \Sigma^{+}_{R} \\\Sigma^{-}_{R} & -\Sigma^0_{R}/\sqrt{2}\end{array}\right) \equiv \frac{\Sigma^i_{R}\sigma^i}{\sqrt{2}},
\end{eqnarray}  
where $\sigma^i$ are the Pauli matrices and $\Sigma^{\pm}_{R} = (\Sigma^1_{R}\mp i\Sigma^2_{R})/\sqrt{2}$. One can 
also write the charge conjugated form of $\Sigma_{R}$,  
\begin{eqnarray}
\Sigma^c_{R} = \left(\begin{array}{cc}\Sigma^{0c}_{R}/\sqrt{2} & \Sigma^{-c}_{R} \\\Sigma^{+c}_{R} & -\Sigma^{0c}_{R}/\sqrt{2}\end{array}\right).
\end{eqnarray}
Note that $\Sigma^c_{R}$ are left-handed fields. With the definition $\Sigma \equiv \Sigma_{R} + \Sigma^c_{R}$, one 
find the diagonal neutral fields of $\Sigma$ are Majorana fermions while off-diagonal charged fields are Dirac fermions. 
The general Lagrangian involving $\Sigma$ fields is given by 
\begin{eqnarray}
{\cal L}_{\Sigma} = Tr[\bar{\Sigma}i\slashed{D}\Sigma] - \frac{1}{2}Tr[\bar{\Sigma}M_{\Sigma}\Sigma] - \overline{l_{L\alpha}}\sqrt{2}Y^{\dag}_{\Sigma_{\alpha i}}\Sigma_{i}\tilde{H} - H^{T}\epsilon^{T}\bar{\Sigma}_{i}\sqrt{2}Y_{\Sigma_{\alpha i}}l_{L_{\alpha}},     
\end{eqnarray}
where $\epsilon$ is the anti-symmetric tensor, $Y_{\Sigma}$ is the Dirac-neutrino Yukawa coupling matrix, and 
$M_{\Sigma}$ is the Majorana mass matrix of heavy fermion triplets. After the Higgs develops VEV, the neutral 
lepton mass matrix in the basis of $(\nu_{L}, \Sigma^0)^T$ can be written as 
\begin{eqnarray}
\left(\begin{array}{cc}0 & vY^T_{\Sigma}/2\sqrt{2} \\vY_{\Sigma}/2\sqrt{2} & M_{\Sigma}/2\end{array}\right).
\end{eqnarray}
In analogous to the Type-I seesaw mechanism, the effective light neutrino mass matrix $m_{\nu_{\alpha\beta}}$ is 
obtained with the substitutions $Y_{\nu} \rightarrow Y_{\Sigma}$ and $M_{R} \rightarrow M_{\Sigma}$. Besides the Yukawa 
interaction, the triplets in type-III seesaw model have gauge interactions and will contribute to the running of gauge coupling, 
additionally, changing the running of $\lambda$ indirectly. 
  
\subsection{The darkon and minimal dark matter}
There are convincing evidences from cosmology and astrophysics that $23\%$ of the energy density of the Universe 
is provided by dark matter~\cite{pdg}. The quest for the nature of the missing component boost the investigations 
on both theoretical and experimental sides. One of the popular DM candidates is the weakly interacting massive 
particle (WIMP) which is stable on cosmological time scale and a thermal relic of the Big Bang. Typically, one of the 
simplest way to justify the stability of DM is to impose a $Z_{2}$ parity symmetry, for example the R-parity in supersymmetry. 
In order to study the vacuum stability bounds of Higgs sector in the extension of SM including neutrino and DM sectors, 
we choose two models of DM which minimally change the SM Higgs potential and keep the direct relation between 
Higgs quartic-coupling $\lambda$ and the electroweak vacuum $v$.    

\begin{figure}[ht]
  \centering
\includegraphics[width=0.65\textwidth]{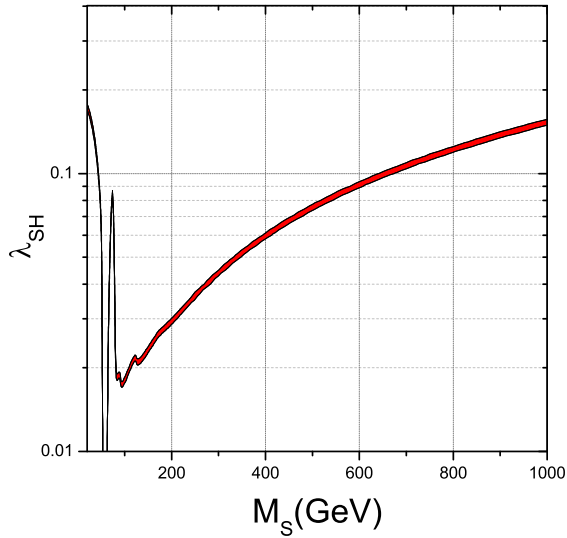}
\caption{Constraint on $M_S$ and $\lambda_{SH}$ from relic density, $\Omega_{\textrm{cdm}}h^{2}=0.111\pm0.006$ \cite{pdg}. 
Recent study \cite{Djouadi:2011aa} shows that darkon with $M_S>80$ GeV or at the resonant region around 62 GeV 
is still allowed by both invisible decay and XENON100 constraints.}
\label{fig:relic_density_darkon}
\end{figure}

\emph{The darkon}: The SM with an additional real singlet scalar $S$~\cite{Silveira:1985rk}, called darkon. 
To stabilize the darkon being a good WIMP candidate, one introduce a discrete $Z_{2}$ symmetry into the model. 
$S$ is odd under the $Z_{2}$ transformation 
while all the SM particles are even. The Lagrangian involving $S$ reads 
\begin{eqnarray}
{\cal L}_{S} = \frac{1}{2}\partial_{\mu}S\partial^{\mu}S - \frac{m^2_{0}}{2}S^2 - \frac{\lambda_{S}}{4}S^4 
- \lambda_{SH}S^2H^{\dag}H. 
\end{eqnarray}  
After electroweak symmetry breaking, the Lagrangian ${\cal L}_{S}$ becomes  
\begin{eqnarray}
{\cal L}_{S} = \frac{1}{2}\partial_{\mu}S\partial^{\mu}S - \frac{m^2_{0} + \lambda_{SH}v^2}{2}S^2 
- \frac{\lambda_{SH}}{2}S^2h^2 - \lambda_{SH}vS^2h,
\end{eqnarray}  
where we identify $h$ the physical Higgs boson. This is one kind of Higgs-portal models, the interactions 
between the darkon and Higgs boson play an important role in determining the relic abundance of 
DM~\cite{McDonald:1993ex,Burgess:2000yq} (also see Fig.~\ref{fig:relic_density_darkon}) and 
the invisible decay width of Higgs boson~\cite{He:2008qm,Cai:2011kb}. Recent study \cite{Djouadi:2011aa} 
shows that darkon with $M_S>80$ GeV or at the resonant region around 62 GeV is still allowed by both invisible 
decay and XENON100 constraints \cite{Aprile:2011ts}.  

\emph{The minimal dark matter}: The authors of \cite{Cirelli:2005uq} first proposed the idea of "minimal dark matter". 
They take the assumption of the existence of only SM gauge symmetry as a guidance and do not introduce extra 
features (some discrete symmetry, for example) to guarantee the stabilization of DM candidate. In \cite{Cirelli:2005uq} 
they found an extra electroweak multiplet field $\chi$ which carries minimal SM quantum numbers and has one lightest 
neutral component after quantum corrections, and furthermore, no operator can be written for the $\chi$ decay such 
that neutral component of $\chi$ is a viable DM candidate and stable particle. MDM also satisfy all the direct/indirect 
dark matter search experiments. It was found that the minimal construction of such field (MDM) $\chi$ is 
a "\emph{fermionic $SU(2)_{L}$ quintuplet with hypercharge $Y = 0$}". $\chi$ is a vector field with respect 
to $SU(2)_{L}$ symmetry and hence the theory is anomaly free. Since $\chi$ transforms as $\textbf{5}$ representation 
under $SU(2)_{L}$, its neutral component only interacts with the SM particles via Weak and gravity forces.  
Therefore, it is a WIMP candidate, and has no free parameter. The mass of MDM is $9.6$ TeV which is determined 
by the relic density of DM~\cite{Cirelli:2009uv}.  

It would be interesting to study the impact on the vacuum stability bounds of the Higgs boson within the above 
frameworks of the SM extensions including neutrinos and DM. We study the vacuum stability and perturbativity 
bounds in the combined four scenarios of SM extension: (1) Type-I seesaw + darkon, (2) Type-I seesaw + MDM, 
(3) Type-III seesaw + darkon, and (4) Type-III seesaw + MDM in the following section. 

\section{Vacuum stability and perturbativity}
\begin{figure}
\centering
\includegraphics[width=0.45\textwidth]{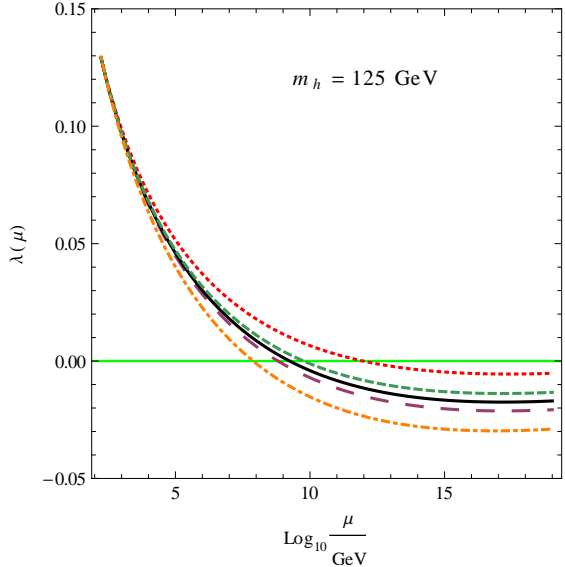}
\caption{Running of $\lambda(\mu)$ in standard model. Black solid line is plotted with $m_t=173.2$ GeV and $\alpha_s(M_Z)=0.1184$. Dotted and Dot-Dashed lines shows the effects of changing mass of $m_t$, for $m_t=171.4$ GeV and $m_t=175.0$, respectively. Two lines, closer to the solid one, display the effects of different value for $\alpha_s(M_Z)$, dashed one for $\alpha_s(M_Z)=0.1198$ and long dashed for $\alpha_s(M_Z)=0.1170$. }
\label{fig:SM_Stability}
\end{figure}

We begin with the brief introduction of vacuum stability and perturbativity in the SM. It is known 
(see~\cite{Cabibbo:1979ay,Lindner:1985uk,Sher:1988mj,Lindner:1988ww, Arnold:1991cv, Altarelli:1994rb, 
Casas:1996aq, Schrempp:1996fb, Isidori:2001bm, Espinosa:2007qp} and refs therein) that the SM Higgs potential 
is modified by high order quantum corrections. For Higgs quartic coupling $\lambda$, its $\beta$ function, $\beta_{\lambda}$, 
receives positive contributions from scalars and gauge bosons and negative contributions from fermions. 
Especially, when $\lambda$ is too large as an input at the electroweak scale, the self-interaction will drive 
$\lambda(\mu)$ to a non-perturbative region and may confront the theory with Landau pole or triviality. 
And when $\lambda$ is too small, it will become negative due to the large top quark Yukawa coupling at 
certain energy scale $\Lambda$, the scale corresponds to the instability of the electroweak vacuum and the 
new physics will appear at this scale. With fixed $m_h$, the scale $\Lambda$ is sensitive to the top quark mass 
and strong coupling constant (see Fig.~\ref{fig:SM_Stability}).

Before the LHC's running, with $m_h$ as a free and unknown parameter, some relevant works on vacuum 
stability has been studied in extensions of SM, including the scalar singlet~\cite{Gonderinger:2009jp,Clark:2009dc,Lerner:2009xg}, 
Type-I seesaw~\cite{Casas:1999cd,EliasMiro:2011aa}, Type-II seesaw mechanism~\cite{Gogoladze:2008gf}, 
and Type-III seesaw mechasnism~\cite{Gogoladze:2008ak}. And recent related works with $m_h$ around $125$ GeV are ~\cite{EliasMiro:2011aa} with Type-I seesaw, \cite{Holthausen:2011aa, Xing:2011aa} in SM, and \cite{Kadastik:2011aa} with scalar dark matter. To our knowledge, no combination or comparison 
of both seesaw mechanism and DM extension of SM has been considered. With the suggestive hint at the LHC, in what 
follows, we take the SM Higgs mass $m_{h} = 125$ GeV as an input and analyze the stability in different frameworks. 

Before studying of stability in the combination scenarios of neutrino and DM two sectors, we shall firstly compare the qualitative difference in each sector and show the quantitative effects of individual parameters. Then we show how the $\lambda(\mu)$ is changed with both neutrino and DM extensions.
  
\subsection{SM + Type-I/Type-III seesaw mechanism}
Both the Type-I and Type-III seesaw mechanisms have the new Yukawa couplings 
($Y_{\nu}$ and $Y_{\Sigma}$ respectively) with Higgs bosons. These bring the additional corrections 
to the $\beta$ function of $\lambda$: 
\begin{eqnarray}\label{beta_type1}
\Delta \beta_{\lambda_{\rm{I}}}=\frac{1}{(4\pi)^2}\left[-4\textrm{ Tr}Y_{\nu}Y_{\nu}^{\dagger}Y_{\nu}Y_{\nu}^{\dagger} +4\lambda\textrm{ Tr}Y_{\nu}Y_{\nu}^{\dagger}\right]
\end{eqnarray}
and
\begin{eqnarray}\label{beta_type3}
\Delta \beta_{\lambda_{\rm{III}}}=\frac{1}{(4\pi)^2}\left[-20\textrm{ Tr}Y_{\Sigma}Y_{\Sigma}^{\dagger}Y_{\Sigma}Y_{\Sigma}^{\dagger} + 12\lambda\textrm{ Tr}Y_{\Sigma}Y_{\Sigma}^{\dagger}\right],
\end{eqnarray}
for Type-I seesaw and Type-III seesaw respectively. For simplicity, we study the case that there is a mass 
hierarchy of light neutrino masses ($m_{\nu_{3}}>m_{\nu_{2}}>m_{\nu_{1}}$, for example) due to the hierarchy 
of Yukawa couplings, and hence, the masses of the heavy neutrino sectors are degenerate. In this case, we 
consider only one sizable Yukawa corresponding to one heavy neutrino sector which will affect the RG evolution 
of $\lambda$\footnote{We assume the light neutrino masses are around $\sqrt{\Delta m^2_{\rm{atm}}} \approx 0.05$ eV in our analysis.}. The triplet 
fermions in Type-III seesaw will also change the $SU(2)_L$ gauge coupling $g_2$ RG 
equation with the modification, 
\begin{equation}
\Delta\beta_{g_{2_{\rm{III}}}}=\frac{1}{(4\pi)^2}\frac{4n}{3},
\end{equation}
here $n$ is the generations of heavy triplets, and we assume $n=3$. The runnings of these Yukawa coupling are also taken into account, governed by their $\beta$ functions in the Appendix.
\begin{figure}[t]
\includegraphics[scale=0.7]{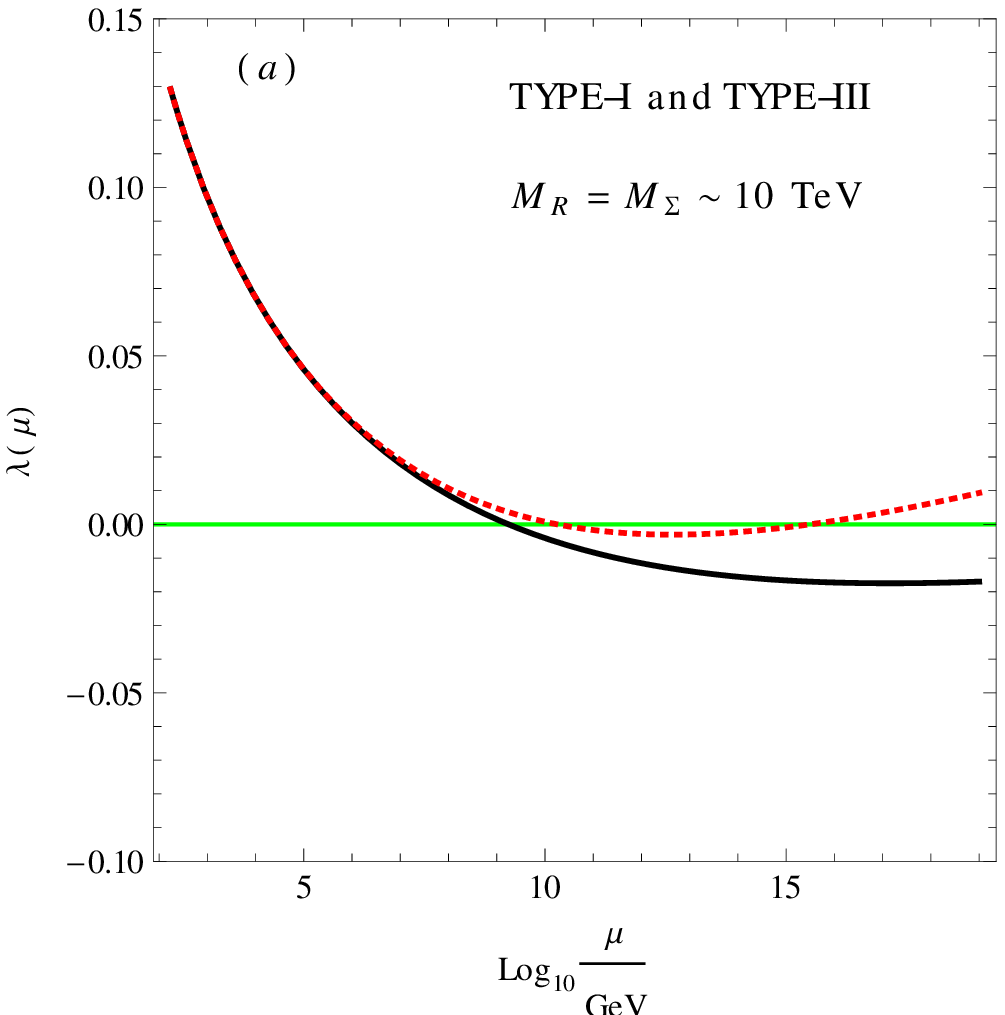}
\includegraphics[scale=0.7]{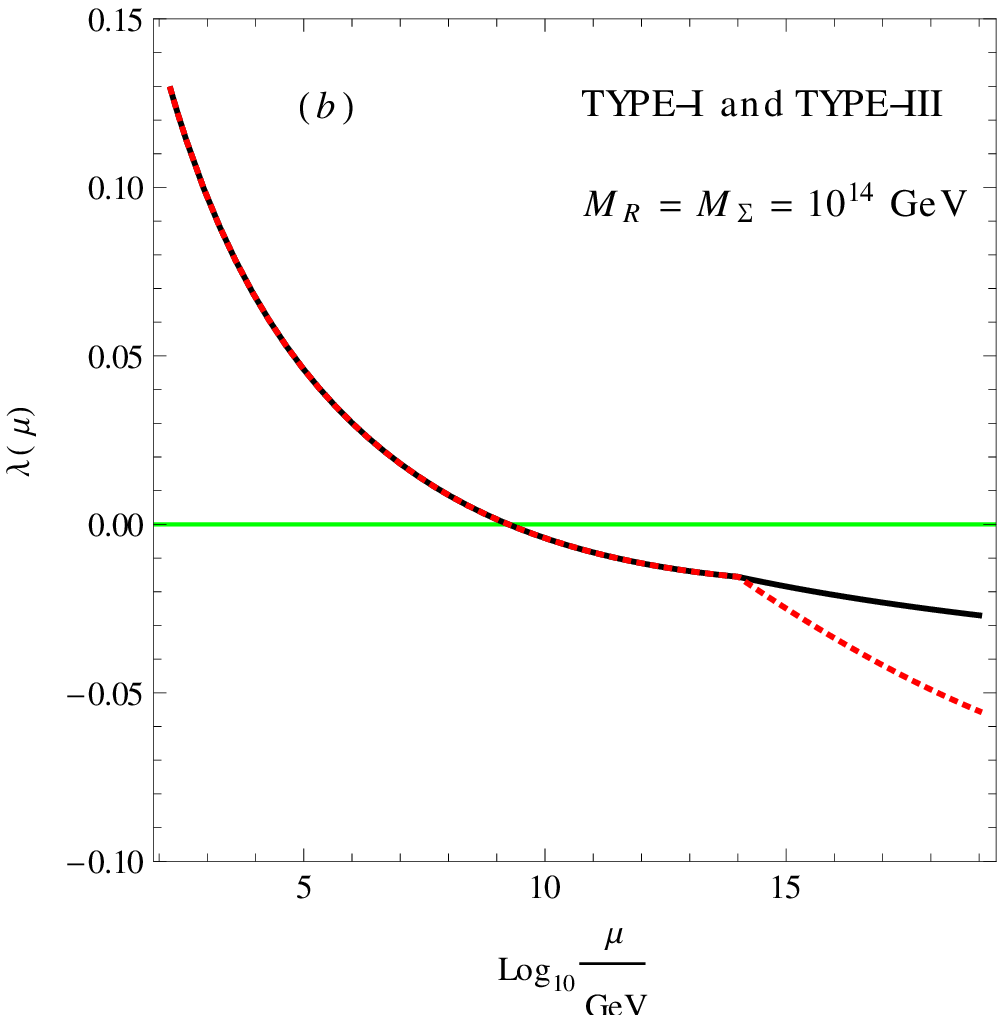}
\caption{Comparison between type-I and type-III seesaw mechanisms for different mass scales of heavy neutrino. 
(a) $M_R=M_{\sigma}=10$ TeV, the black solid line shows the running of $\lambda(\mu)$ in type-I seesaw and the red dotted one for type-III seesaw. 
(b) $M_R=M_{\sigma}=10^{14}$ GeV, the black solid line show the running of $\lambda(\mu)$ in type-I seesaw and the red dotted one for type-III seesaw. The threshold effect can be easily seen at $\mu = 10^{14}$ GeV.}
\label{fig:type_13}
\end{figure}

In Fig.~\ref{fig:type_13} we compare the RG evolutions in the cases of Type-I and Type-III seesaw mechanisms extension with different mass scales of heavy fermions. As shown in the figures, when the mass scale of the heavy neutrinos is smaller than $10^{14}$~GeV, these neutrion Yukawa couplings is small and can be ignored. The deviation between Type-I and Type-III seesaw mechanisms is caused dominantly by the gauge coupling running when $M_R=M_{\Sigma}=10$ TeV in $(a)$ of Fig.~\ref{fig:type_13}, and by fermion degrees of freedom and when $M_R=M_{\Sigma}=10^{14}$ GeV in $(b)$ of Fig.~\ref{fig:type_13}.    

\begin{figure}
\centering
\includegraphics[width=0.4\textwidth]{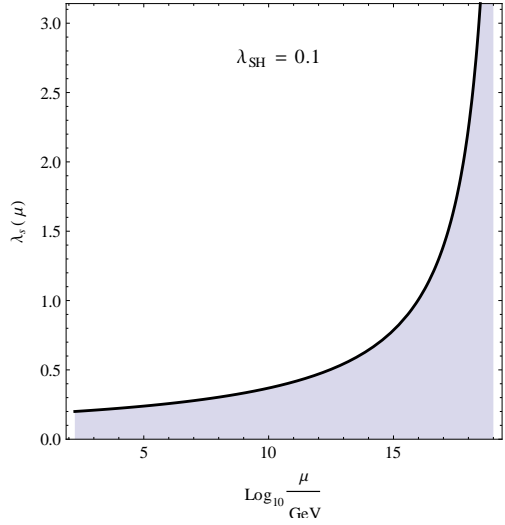}
\caption{Perturbativity bound for $\lambda_{S}$. From the constraint of relic density, when $M_S \sim 1$ TeV, $\lambda_{SH}\sim \mathcal{O}(0.1)$ shown in Fig. (\ref{fig:relic_density_darkon}). If we further require that $\lambda_S (10^{19})<\pi$ for perturbativity, then $\lambda_S (M_S)<0.2$ at the input scale. }
\label{fig:triviality_lambda_S}
\end{figure}

\subsection{SM + darkon/MDM}
Now we consider the cases of SM extension of dark matter sector. For darkon model, we first scan the parameter space 
which is allowed by various constraints. The relic abundance for DM density is shown previously in Fig.~\ref{fig:relic_density_darkon}. Since $\lambda_{S}$ always receives positive contributions from Higgs boson, we do not have to worry about the stability of $S$ at large field values. Instead, we plot the perturbative bounds for $\lambda_{S}$ in Fig.~\ref{fig:triviality_lambda_S} and find that in the range of $0<\lambda_{S}<0.2$ one can satisfy the perturbative bounds up to $M_{\rm{Pl}}$ with $\lambda_{SH} = 0.1$ which provides the strongest constraint as shown in Fig.~\ref{fig:relic_density_darkon}. 

The scalar singlet will modify the $\beta_\lambda$ as  
\begin{equation}
\Delta  \beta_{\lambda_{\rm{darkon}}} = \frac{1}{(4\pi)^2}2 \lambda_{SH}^2,
\end{equation}
accompanied by $\beta$ functions for darkon sector in the Appendix. We also plot in Fig.~\ref{fig:constraint_darkon}($(a)$,$(b)$) to show the sensitivity of $\lambda$ RG running to $\lambda_{S}$ and $\lambda_{SH}$ respectively, and Fig.~\ref{fig:constraint_darkon}$(c)$ implies the stabilization holds as long as $\lambda_{SH}\gtrsim0.031$ for $\lambda_{S}=0.1$. 

\begin{figure}[ht]
\includegraphics[scale=0.75]{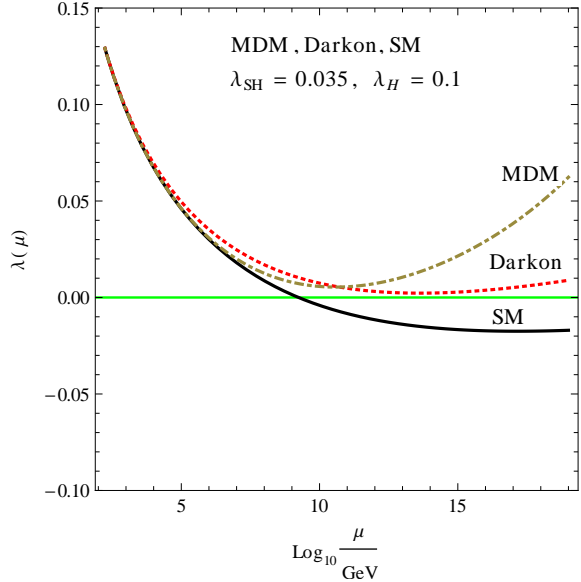}
\caption{Comparison between MDM and darkon's effects on the running of $\lambda(\mu)$. Unlike the Darkon case, in MDM extension there is no free parameter after the mass is fixed by relic density and $\lambda(\mu)>0$ is valid up to Planck scale.}
\label{fig:MDM_Darkon_SM}
\end{figure}

Rather than contributing $\beta_\lambda$ directly, the minimal dark matter will change the RG evolution of $SU(2)_L$ gauge coupling $g_2$ with 
\begin{equation}\label{beta_MDM}
\Delta\beta_{g_{2_{\rm{MDM}}}}=\frac{1}{(4\pi)^2}\frac{20}{3},
\end{equation}
and as a result, gives a positive contribution to $\beta_\lambda$ indirectly. We note that there is no free parameter in the case of MDM, therefore we plot in Fig.~\ref{fig:MDM_Darkon_SM} to show the differences between the SM and the extension of darkon and MDM. The positive contribution to $\lambda(\mu)$ is completely due to the growing of gauge coupling $g_{2}(\mu)$ in MDM scenario. In the case of darkon, the raise of $\lambda(\mu)$ is more for larger $\lambda_{SH}$ and $\lambda_{S}$ while the value of $\lambda_{SH}$ has to be consistent with the relic abundance of DM. 

\begin{figure}
\includegraphics[width=0.328\textwidth]{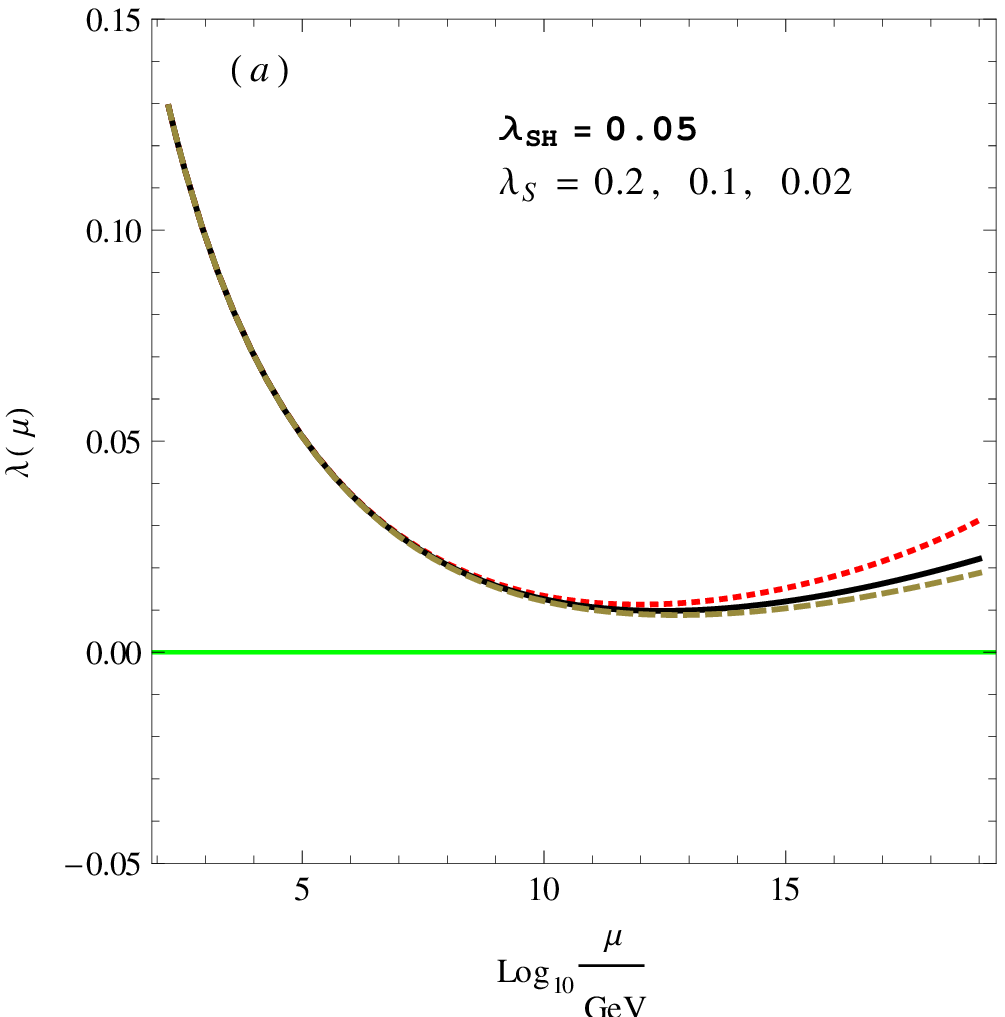}
\includegraphics[width=0.328\textwidth]{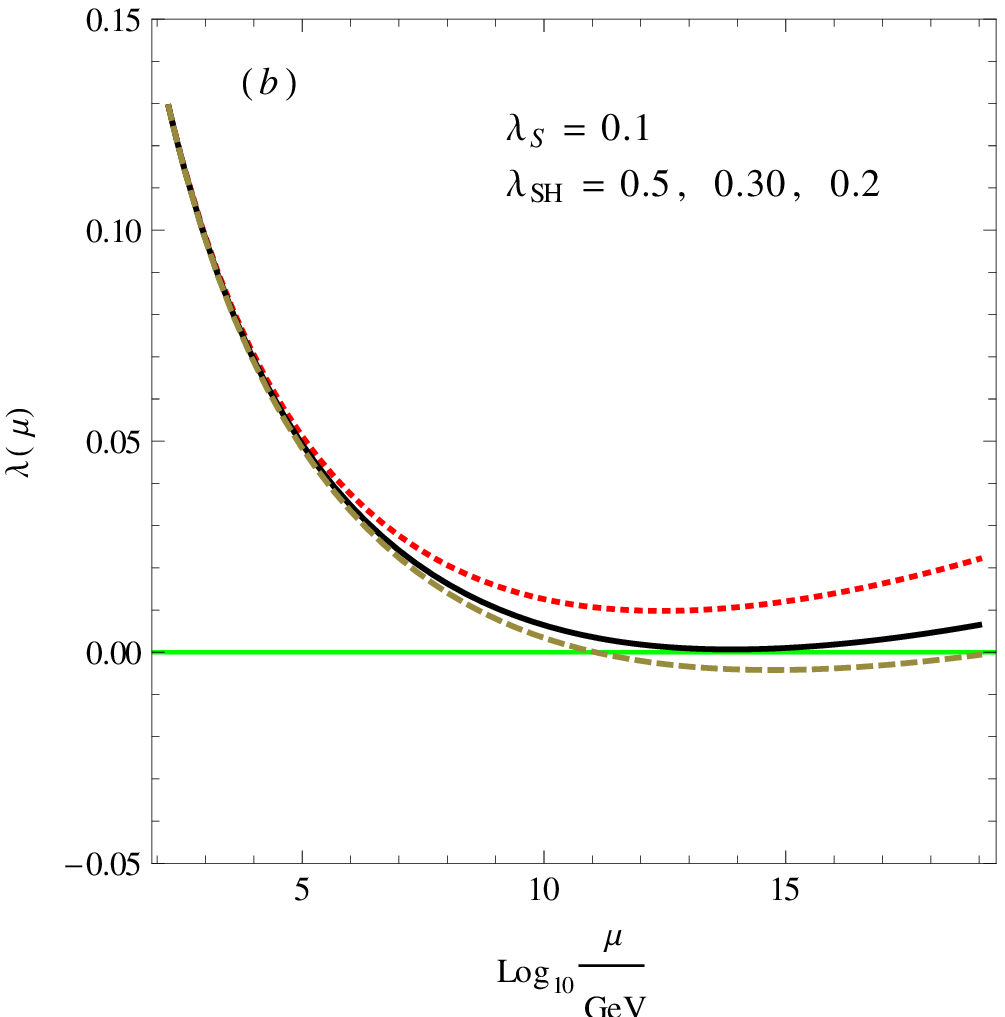}
\includegraphics[width=0.328\textwidth]{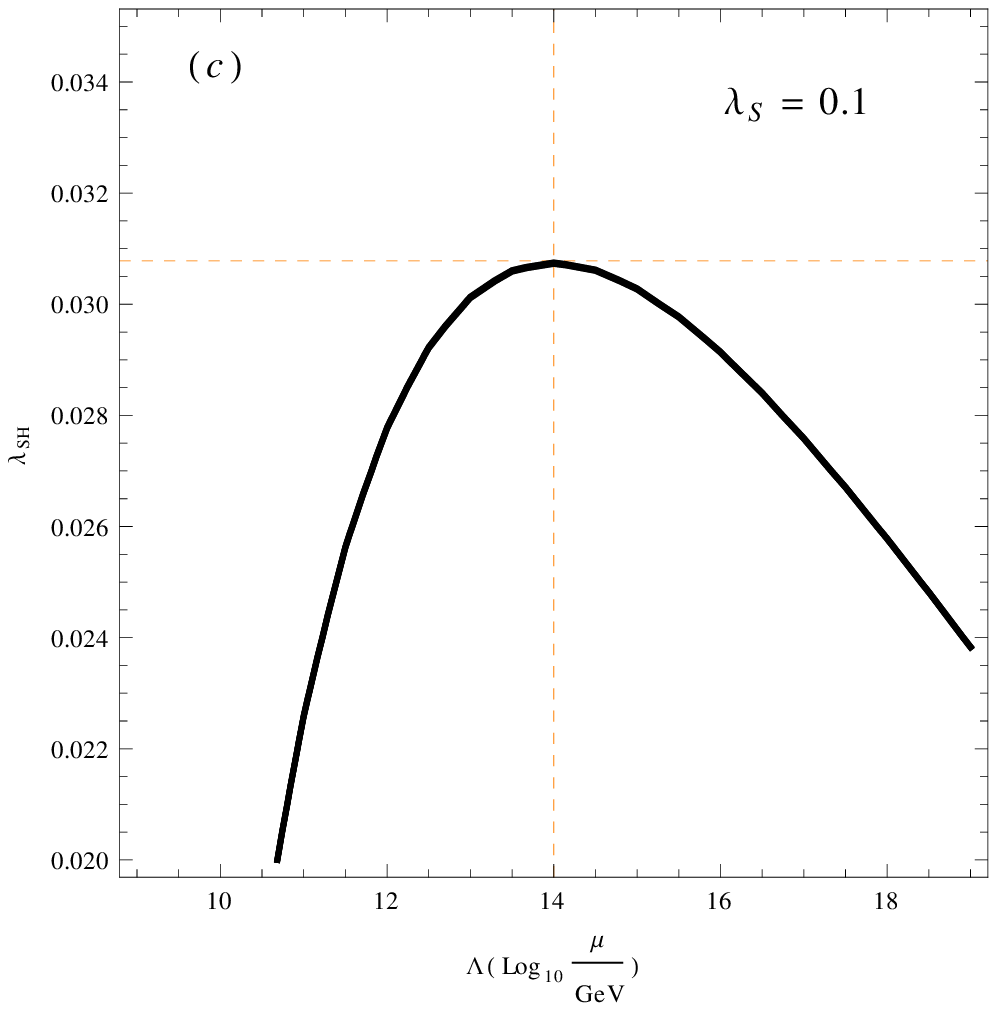}
\caption{Effects on running of $\lambda$ when varying $\lambda _S$ and $\lambda _{SH}$, respectively. 
$(a)$ Dotted, solid and dashed lines correspond to $\lambda _S=0.2, 0.1 \textrm{ and } 0.02$, respectively.
$(b)$ Dotted, solid and dashed lines correspond to $\lambda _{SH}=0.5, 0.3 \textrm{ and } 0.2$, respectively.
$(c)$ show $\Lambda$ at which $\lambda(\Lambda)=0$ for different $\lambda _{SH}$, $\lambda _S$ is fixed here as an example. When $\lambda _S > 0.031$, stability can be preserved up to Planck scale.}
\label{fig:constraint_darkon}
\end{figure}

Confronting the unsolved puzzles of neutrino mass and dark matter in the SM, we study the extensions of SM 
on the Type-I(III) seesaw and darkon (MDM) with the four combinations in the following. 

\begin{figure}
\includegraphics[scale=0.75]{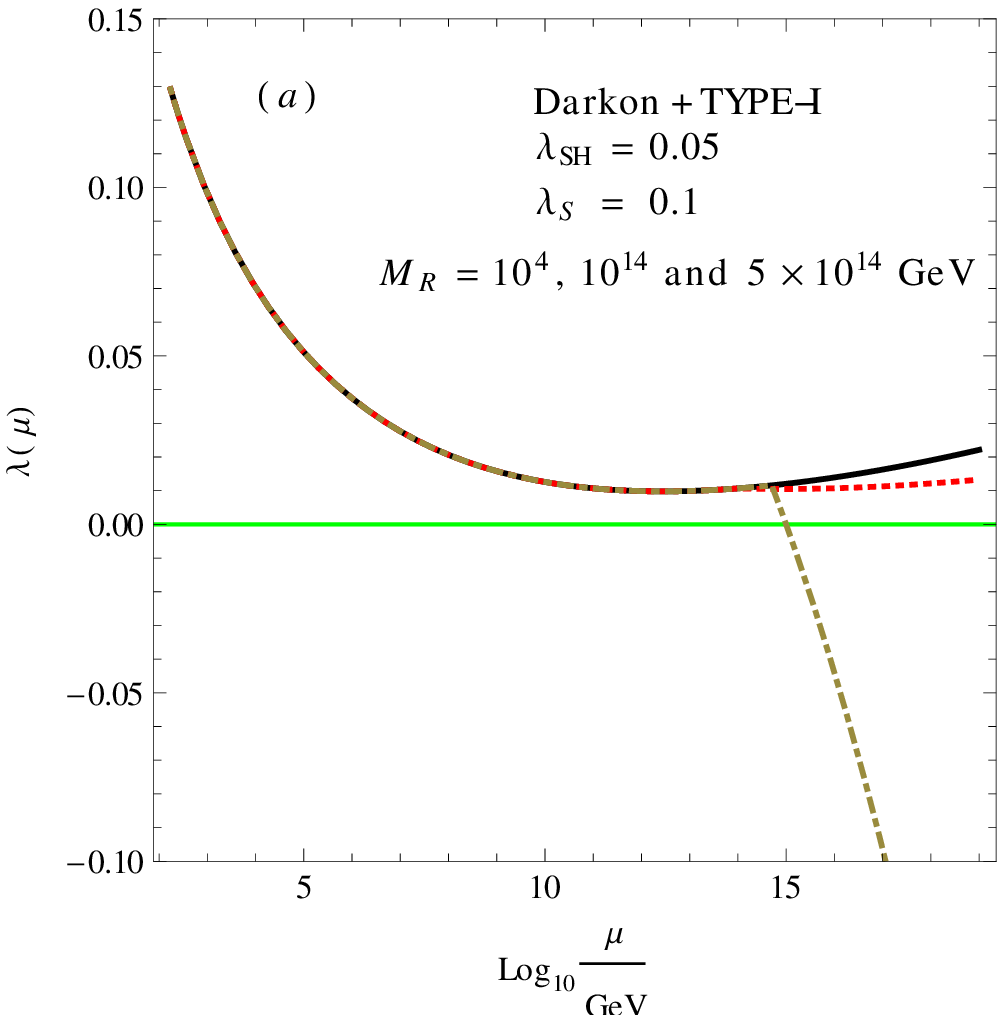}
\includegraphics[scale=0.75]{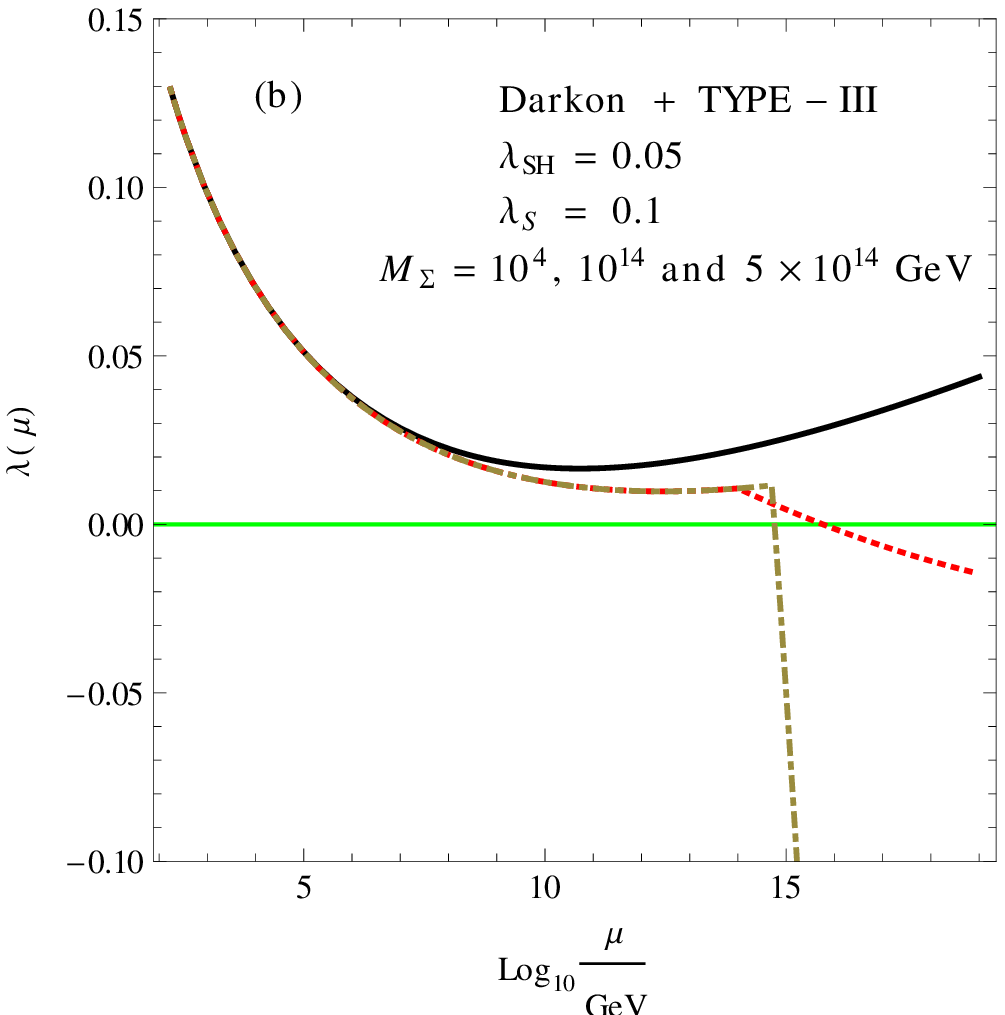}
\caption{Combination of Darkon and neutrino in type-I and type-III seesaw mechanisms. The qualitative features of the running of $\lambda(\mu)$ are the same in $(a)$ and $(b)$. And quantitative differences results from the neutrino's contribution to $\beta_{\lambda}$ in two scenarios. In both $(a)$ and $(b)$, solid, dotted and dot-dashed lines represent the effects from different mass scales of heavy neutrinos, $10^{4}$, $10^{14}$ and $5\times 10^{14}$ GeV, respectively.}
\label{fig:Darkon_Type_13}
\end{figure}

\underline{\emph{SM + Type-I/Type-III seesaw + darkon}}: 

The figure is plotted in Fig.~\ref{fig:Darkon_Type_13} which shows that when Yukawa couplings are smaller 
than ${\cal O}(0.1)$, both Type-I and Type-III seesaw are irrelevant due to the forth power suppression 
of $Y_{\nu,\Sigma}$ in $\beta_{\lambda_{\rm{I,III}}}$. In other words, the running of $\lambda(\mu)$ is 
sensitive to the scales of heavy neutrinos as we see in Fig.~\ref{fig:Darkon_Type_13}. When the scales 
of the heavy neutrino sectors are large (${\cal O}(10^{14-15})$~GeV), the Yukawa couplings are sizable 
and we see the instability of electroweak vacuum occurs around the new neutrino scales. The drop of 
$\lambda(\mu)$ is sharper for the Type-III seesaw due to more degrees of freedom (triplet fermion) involved 
in the quartic-coupling box diagram (see Eqs.~(\ref{beta_type1}),(\ref{beta_type3})). The effects due to the 
growing of gauge coupling $g_{2}$ is relatively weaker in these scenarios.    

\begin{figure}
\includegraphics[scale=0.75]{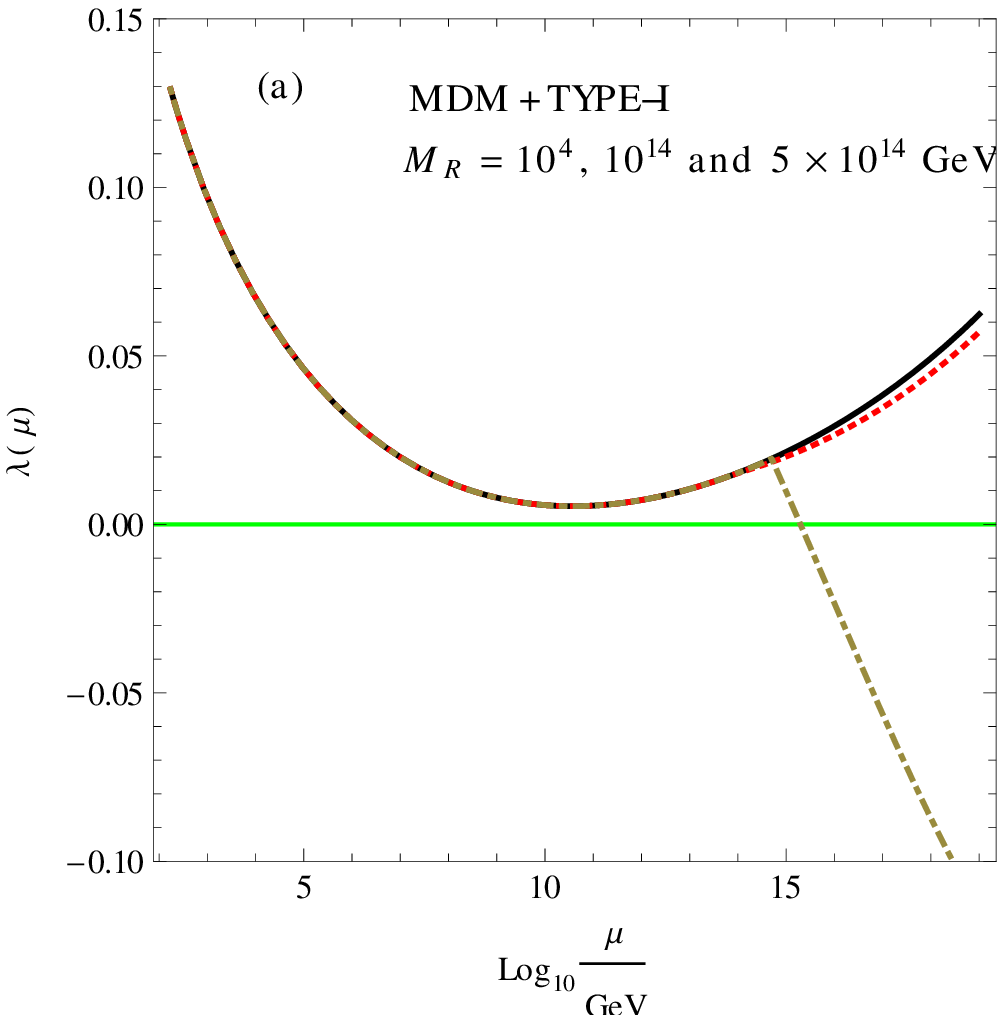}
\includegraphics[scale=0.75]{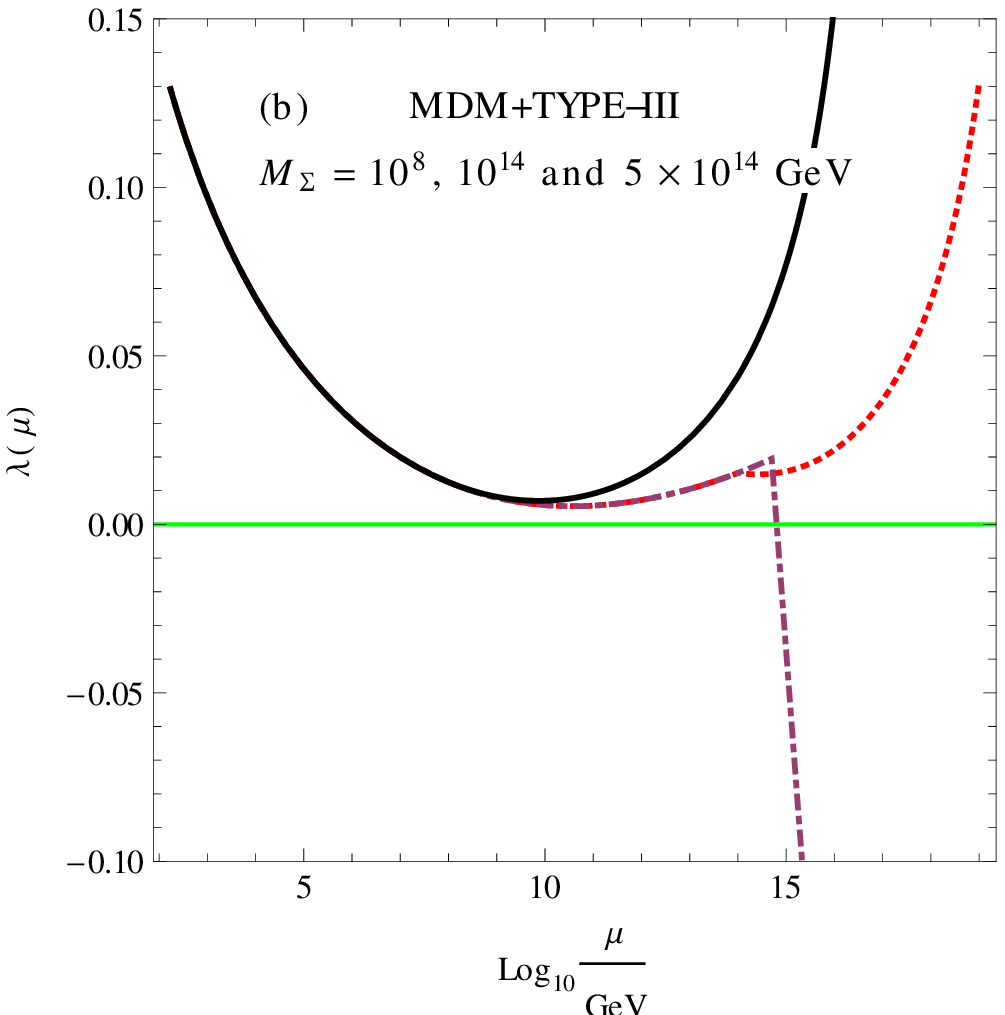}
\caption{Combination of MDM and neutrino in type-I and type-III seesaw mechanisms. In both $(a)$ and $(b)$, solid, dotted and dot-dashed lines represent the effects from different mass scales of heavy neutrinos, $10^{4}$, $10^{14}$ and $5\times 10^{14}$ GeV, respectively.}
\label{fig:MDM_Type13}
\end{figure}

\underline{\emph{SM + Type-I/Type-III seesaw + MDM}}:

In these cases the effects of gauge coupling $g_{2}(\mu)$ running are important to $\lambda(\mu)$. Since 
the mass of MDM is 9.6 TeV fixed by relic density of DM \cite{Cirelli:2009uv} and the $g_{2}(\mu)$ starts to receive the additional contribution (Eq.~(\ref{beta_MDM})) from MDM at this scale. This will cause the raise of $\lambda(\mu)$ via 
the growing of $g_{2}(\mu)$ to compensate for the negative contributions from Yukawa couplings and avoid the 
instability at high scales. From Fig.~\ref{fig:MDM_Type13} we see how the $g_{2}(\mu)$ modify the evolutions of 
$\lambda(\mu)$ in Type-I/Type-III plus MDM frameworks. One should notice that one of the conditions of minimal 
dark matter is the perturbativity of $g_{2}$ to be valid up to Planck scale. If the three generation triplet fermions of 
Type-III seesaw are lighter than $10^{8}$~GeV, the perturbativity of $g_{2}(\mu)$ will blow up at the scale 
around ${\cal O}(10^{15})$~GeV. The validity of fermionic quintuplet as minimal dark matter may be quested in this 
extended scenario. However, the mass spectrum of and the generation numbers of $\Sigma_{i}$ will alter the 
$g_{2}$ evolution and relax the perturbativity bound up to $M_{\rm{Pl}}$. We will not address this discussion in this paper.           

\begin{figure}
\includegraphics[scale=0.75]{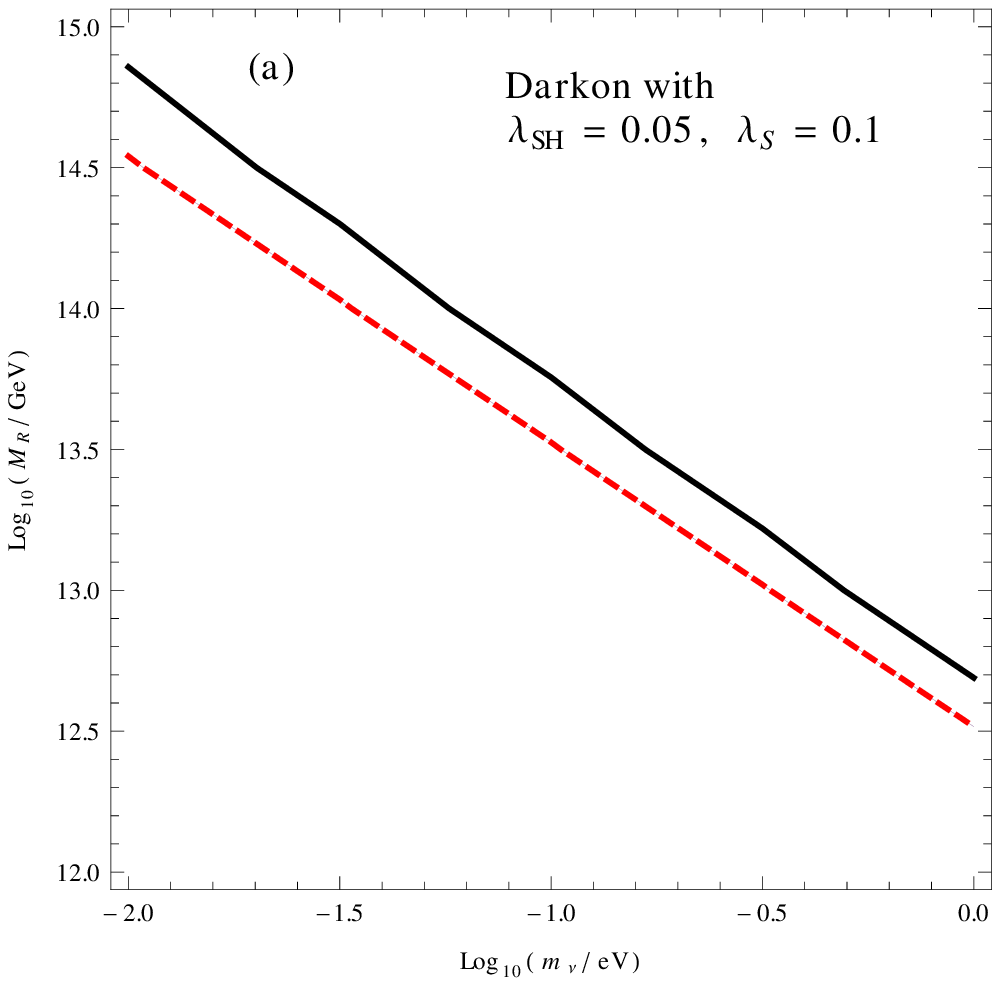}
\includegraphics[scale=0.75]{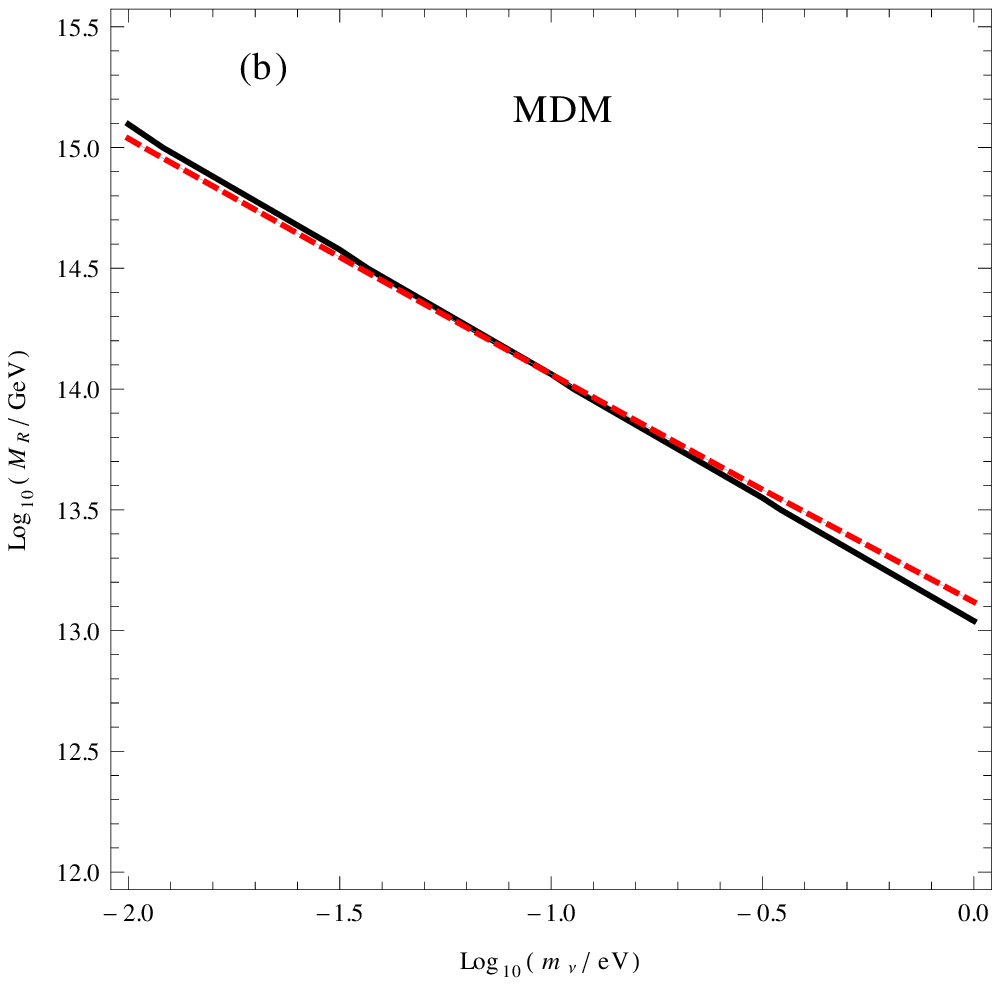}
\caption{Constraint on the neutrino sector with the requirement $\lambda(M_{Pl})>0$. In both $(a)$ and $(b)$, solid  and dashed lines display the contours at $\lambda(M_{Pl})= 0 $ for type-I and type-III seesaw mechanism, respectively. Regions under the line satisfy $\lambda(M_{Pl})>0$.}
\label{fig:Neutrino_DM}
\end{figure}

At last, we show how stability can constrain the mass scales in the neutrino sector in Fig.~\ref{fig:Neutrino_DM}. 
If there is no new physics other than dark matter and neutrino up the Planck scale, $\lambda(M_{\rm{Pl}})>0$ can 
put bounds on neutrino part  with fixed parameters in DM sector. In $(a)$ of Fig.~\ref{fig:Neutrino_DM}, the 
contours $\lambda(M_{\rm{Pl}})=0$ are plotted in the framework of Darkon$+$Type-I/III. Here bounds on 
Type-III seesaw are more stringent because of the magnitude of its contribution to $\beta_{\lambda}$ as also 
suggested in Fig.~\ref{fig:Darkon_Type_13}. With the combination of MDM$+$Type-I/III, $(b)$ of 
Fig.~\ref{fig:Neutrino_DM} shows the individual bounds. Contour lines are closer than those in $(a)$, a feature of 
the competition between gauge coupling and Yukawa's contributions.

\section{Conclusions}
We study the stability of the electroweak vacuum within the SM and its extensions of including the Type-I(III) seesaw 
mechanisms and dark matter candidates (darkon and minimal dark matter) and also discussed the stability bounds on 
the scales of neutrino sector. Motivated by the possible  evidence of the SM Higgs at the LHC, we apply the signal 
of $m_{h} = 125$~GeV as the input and investigate the RG evolutions of Higgs quartic-coupling $\lambda$ in four 
scenarios:(1) Type-I seesaw + darkon, (2) Type-I seesaw + MDM, (3) Type-III seesaw + darkon, and (4) Type-III seesaw + MDM. 
The scalar singlet contributes positively to the $\beta$ function of $\lambda$ to stabilize the SM vacuum up to high scale. 
However, similar to the top quark mass in the SM, the results depend sensitively on the Yukawa couplings $Y_{\nu,\Sigma}$ 
between the new neutrino sectors and SM Higgs. If $Y_{\nu,\Sigma}\sim{\cal O}(1)$, it will bring the instability at the heavy 
neutrino mass scale. Additionally, minimal dark matter and fermionic triplets of Type-III seesaw mechanism transform 
nontrivially under the $SU(2)_{L}$ gauge group, and hence, will boost the growth of gauge coupling $g_{2}(\mu)$ 
running. As a result, they will bring the positive contributions to save the stability of Higgs vacuum.

\acknowledgments
This work is supported by National Center for Theoretical Sciences, Taiwan. 

\appendix
\section{Renormalization Group(RG) Equations}
\label{App}

In this appendix we list the relevant input and RG equations in our numerical evaluation. 
\begin{eqnarray*}
&&m_h=125 \textrm{ GeV},\;m_t=173.2\pm 0.9 \textrm{ GeV},\;M_Z=91.188\textrm{ GeV},\\
&&\alpha_s(M_Z)=0.1184\pm 0.0007,\;\alpha(M_Z)=1/127.926, \;\textrm{sin}{}^2\theta(M_Z)=0.2312.
\end{eqnarray*}
And $\lambda$ and $y_t$ are related with $m_h$ and $m_t$ by matching condition \cite{Hambye:1996wb}, respectively.

{\it Standard Model} : For the standard model, we collect RG equations at two-loop order for Higgs quartic coupling, top-quark Yukawa coupling and the gauge couplings,  $\lambda$, $y_t$ and $g_i$ \cite{Cheng:1973nv, Machacek:1983tz,Machacek:1983fi,Machacek:1984zw, Arason:1991ic}. For the Higgs quartic coupling we have
\begin{eqnarray}
\beta_\lambda&=&
 {1\over(4\pi)^2}  \left[24  \lambda ^2-6 y_t^4+\frac{3}{8} \left(2 g^4+\left(g^2+g'^2\right)^2\right)+\left(-9 g^2-3
   g'^2+12 y_t^2\right) \lambda \right]\nonumber\\
   &+&{1\over (4\pi)^4}
  \Bigg{[}\frac{1}{48} \left(915 g^6-289 g^4 g'^2-559 g^2 g'^4-379 g'^6\right)+30
    y_t^6-y_t^4 \left(\frac{8 g'^2}{3}+32 g_s^2+3 \lambda
   \right)\nonumber\\
   &+& \lambda  \left(-\frac{73}{8} g^4+\frac{39}{4} g^2 g'^2+\frac{629
   }{24} g'^4+108  g^2  \lambda +36 g'^2 \lambda -312
    \lambda ^2\right)\nonumber\\
   &+& y_t^2 \left(-\frac{9}{4} g^4+\frac{21}{2} g^2
   g'^2-\frac{19}{4}g'^4+ \lambda  \left(\frac{45}{2}g^2+\frac{85
   }{6}g'^2+80 g_s^2-144  \lambda \right)\right)\Bigg{]}.
 \end{eqnarray}
For the top-quark Yukawa coupling we have
\begin{eqnarray}
\beta_{y_t} &=&
 {y_t\over(4\pi)^2} \left[\frac{9}{2} y_t^2 -\frac{9}{4} g^2-\frac{17
   }{12}g'^2-8 g_s^2\right]
+{y_t\over(4\pi)^4}
   \Bigg{[}-\frac{23}{4} g^4-\frac{3}{4} g^2
   g'^2+\frac{1187 }{216}g'^4+9 g^2
   g_s^2 \nonumber\\
 & + &  \frac{19}{9} g'^2 g_s^2-108
   g_s^4+\left(\frac{225}{16}g^2+\frac{131 }{16}g'^2+36 g_s^2\right) 
   y_t^2 + 6 \left(-2  y_t^4-2
    y_t^2 \lambda + \lambda   ^2\right)\Bigg{]}.
   \label{betayt}\end{eqnarray}
And for the gauge couplings $g_i=\{g',g,g_s\}$, we have
\begin{eqnarray}
\beta_{g_i} & = &{1\over(4\pi)^2}g_i^3 b_i+{1\over(4\pi)^4}g_i^3\left[\sum_{j=1}^3 c_{ij}g_j^2-d_i y_t^2\right],
\end{eqnarray}
with
\begin{equation}
b=(41/6,-19/6,-7),\quad
c=\left(
\begin{array}{ccc}
199/18 & 9/2 & 44/3 \\
3/2 & 35/6 & 12 \\
11/6 & 9/2 & -26
\end{array}\right),\quad
d=(17/6,3/2,2).
\end{equation}

{\it Darkon} : In evaluating the effects from darkon's contribution, $\beta$ functions for darkon \cite{Gonderinger:2009jp} itself are also considered. We have at one-loop order
\begin{equation}
\beta_{\lambda_{SH}}={1\over(4\pi)^2}\left[ 8\lambda_{SH}^2 + 6\lambda_{SH}\lambda_{S} - \frac{9}{2}\lambda_{SH}g^2 - \frac{3}{2}\lambda_{SH}{g'}^2 + 6\lambda_{SH}y_t^2 + 2\lambda_{SH}\lambda\right],\label{eq:a2_beta_function}
\end{equation}
and
\begin{equation}
\beta_{\lambda_{S}}={1\over(4\pi)^2}\left[ 8\lambda_{SH}^2 + 18\lambda_{S}^2\right],
\end{equation}
for the running of $\lambda_{SH}$ and $\lambda_{S}$, respectively.

{\it Neutrino }: The running of Yukawa coupling for the neutrino sector is governed by  \cite{Casas:1999cd,Gogoladze:2008ak,Chakrabortty:2008zh}
\begin{equation}
\mu\frac{d}{d\mu}\left(Y^\dagger_\nu Y_\nu\right)=\frac{1}{(4\pi)^2}Y^\dagger_\nu Y_\nu\left[6y_t^2 + 2\textrm{ Tr}\left(Y^\dagger_\nu Y_\nu\right) - \left( \frac{9}{10}g^2_1 + \frac{9}{10}g^2_2 \right)+3 Y^\dagger_\nu Y_\nu \right]
\end{equation}
for Type-I seesaw and
\begin{equation}
\mu\frac{d}{d\mu}\left(Y^\dagger_\nu Y_\nu\right)=\frac{1}{(4\pi)^2}Y^\dagger_\nu Y_\nu\left[6y_t^2 + 3\textrm{ Tr}\left(Y^\dagger_\nu Y_\nu\right) - \left( \frac{9}{10}g^2_1 + \frac{33}{2}g^2_2 \right)+\frac{5}{2} Y^\dagger_\nu Y_\nu \right]
\end{equation}
for Type-III seesaw mechanisms at one-loop order. When neutrino Yukawa couplings  are not ignorable, they can change the $\beta$-function of top-quark Yukawa by
\begin{equation}
\Delta \beta_{y_t}=\frac{1}{(4\pi)^2}\left[\textrm{ Tr}\left(Y^\dagger_\nu Y_\nu\right) \right]
\end{equation}
in Type-I seesaw and 
\begin{equation}
\Delta \beta_{y_t}=\frac{1}{(4\pi)^2}\left[3\textrm{ Tr}\left(Y^\dagger_\nu Y_\nu\right) \right]
\end{equation}
in Type-III seesaw.


\begin{thebibliography}{99}
\bibitem{ATLAS2012} 
  ATLAS Collaboration,
  %``Combined search for the Standard Model Higgs boson using up to 4.9 fb-1 of pp collision data at sqrt(s) = 7 TeV with the ATLAS detector at the LHC,''
  arXiv:1202.1408 [hep-ex].
  
\bibitem{CMS2012} 
  CMS Collaboration,
  %``Combined results of searches for the standard model Higgs boson in pp collisions at sqrt(s) = 7 TeV,''
  arXiv:1202.1488 [hep-ex].

\bibitem{Weinberg:1979sa}
  S.~Weinberg,
  %``Baryon And Lepton Nonconserving Processes,''
  Phys.\ Rev.\ Lett.\  {\bf 43}, 1566 (1979).
  %%CITATION = PRLTA,43,1566;%%
  
    
\bibitem{seesaw} M. Gell-Mann, P. Ramond, and R. Slansky, Complex
Spinors and Unified Theories. In Supergravity, edited by D.Z.
Freedman and P.van Nieuwenhuizen (North Holland, Amsterdam, 1979)
315; E. Witten, {\it Phys. Lett.} {\bf B 91} (1980) 81; B. Kayser,
F. Gibrat-Debu, and F. Perrier, The Physics of the Massive
Neutrinos, World Scientific, Singapore (1989); R.N. Mohapatra {\it
et al.}, {\it Rept. Prog. Phys.} {\bf 70} (2007) 1757.

\bibitem{Foot:1988aq}
  R.~Foot, H.~Lew, X.~G.~He and G.~C.~Joshi,
  %``SEESAW NEUTRINO MASSES INDUCED BY A TRIPLET OF LEPTONS,''
  Z.\ Phys.\  C {\bf 44}, 441 (1989).
  %%CITATION = ZEPYA,C44,441;%%  
  
\bibitem{seesaw2}
  W.~Konetschny and W.~Kummer,
  %``Nonconservation Of Total Lepton Number With Scalar Bosons,''
  Phys.\ Lett.\  B {\bf 70}, 433 (1977); 
  %%CITATION = PHLTA,B70,433;%%   
T.~P.~Cheng and L.~F.~Li,
  %``Neutrino Masses, Mixings And Oscillations In SU(2) X U(1) Models Of
  %Electroweak Interactions,''
  Phys.\ Rev.\  D {\bf 22}, 2860 (1980);
  %%CITATION = PHRVA,D22,2860;%%
  J.~Schechter and J.~W.~F.~Valle,
  %``Neutrino Masses In SU(2) X U(1) Theories,''
  Phys.\ Rev.\  D {\bf 22}, 2227 (1980);
  %%CITATION = PHRVA,D22,2227;%%
   G.~B.~Gelmini and M.~Roncadelli,
  %``Left-Handed Neutrino Mass Scale And Spontaneously Broken Lepton Number,''
  Phys.\ Lett.\  B {\bf 99}, 411 (1981);
  %%CITATION = PHLTA,B99,411;%%
R.~N.~Mohapatra and G.~Senjanovic,
  %``Neutrino Masses And Mixings In Gauge Models With Spontaneous Parity
  %Violation,''
  Phys.\ Rev.\  D {\bf 23}, 165 (1981).
  %%CITATION = PHRVA,D23,165;%%

 
\bibitem{pdg} 
K. Nakamura et al. (Particle Data Group), J. Phys. G 37, 075021 (2010) 
and 2011 partial update for the 2012 edition. 

\bibitem{Silveira:1985rk} 
 V.~Silveira and A.~Zee,
 %``Scalar Phantoms,''
 Phys.\ Lett.\ B {\bf 161}, 136 (1985).
 %%CITATION = PHLTA,B161,136;%%

  
% % Darkon Dark Matter
 %\cite{Silveira:1985rk}

%\cite{McDonald:1993ex}
\bibitem{McDonald:1993ex} 
  J.~McDonald,
  %``Gauge singlet scalars as cold dark matter,''
  Phys.\ Rev.\ D {\bf 50}, 3637 (1994)
  [hep-ph/0702143 [HEP-PH]].
  %%CITATION = HEP-PH/0702143;%%
  
%\cite{Burgess:2000yq}
\bibitem{Burgess:2000yq} 
  C.~P.~Burgess, M.~Pospelov and T.~ter Veldhuis,
  %``The Minimal model of nonbaryonic dark matter: A Singlet scalar,''
  Nucl.\ Phys.\ B {\bf 619}, 709 (2001)
  [hep-ph/0011335].
  %%CITATION = HEP-PH/0011335;%%
  
%\cite{He:2008qm}
\bibitem{He:2008qm} 
  X.~-G.~He, T.~Li, X.~-Q.~Li, J.~Tandean and H.~-C.~Tsai,
  %``Constraints on Scalar Dark Matter from Direct Experimental Searches,''
  Phys.\ Rev.\ D {\bf 79}, 023521 (2009)
  [arXiv:0811.0658 [hep-ph]].
  %%CITATION = ARXIV:0811.0658;%%
%\cite{Cai:2011kb}
\bibitem{Cai:2011kb} 
  Y.~Cai, X.~-G.~He and B.~Ren,
  %``Low Mass Dark Matter and Invisible Higgs Width In Darkon Models,''
  Phys.\ Rev.\ D {\bf 83}, 083524 (2011)
  [arXiv:1102.1522 [hep-ph]].
  %%CITATION = ARXIV:1102.1522;%%

%\cite{Djouadi:2011aa}
\bibitem{Djouadi:2011aa} 
  A.~Djouadi, O.~Lebedev, Y.~Mambrini and J.~Quevillon,
  %``Implications of LHC searches for Higgs--portal dark matter,''
  arXiv:1112.3299 [hep-ph].
  %%CITATION = ARXIV:1112.3299;%%
  
 %\cite{Aprile:2011ts}
 \bibitem{Aprile:2011ts} 
   E.~Aprile {\it et al.}  [XENON100 Collaboration],
   %``Implications on Inelastic Dark Matter from 100 Live Days of XENON100 Data,''
   Phys.\ Rev.\ D {\bf 84}, 061101 (2011)
   [arXiv:1104.3121 [astro-ph.CO]].
  %%CITATION = ARXIV:1104.3121;%%  
  

\bibitem{Cirelli:2005uq} 
  M.~Cirelli, N.~Fornengo and A.~Strumia,
  %``Minimal dark matter,''
  Nucl.\ Phys.\ B {\bf 753}, 178 (2006)
  [hep-ph/0512090].

\bibitem{Cirelli:2009uv} 
  M.~Cirelli and A.~Strumia,
  %``Minimal Dark Matter: Model and results,''
  New J.\ Phys.\  {\bf 11}, 105005 (2009)
  [arXiv:0903.3381 [hep-ph]].

% Stability in standard model
%\cite{Cabibbo:1979ay}
\bibitem{Cabibbo:1979ay} 
  N.~Cabibbo, L.~Maiani, G.~Parisi and R.~Petronzio,
  %``Bounds on the Fermions and Higgs Boson Masses in Grand Unified Theories,''
  Nucl.\ Phys.\ B {\bf 158}, 295 (1979).
  %%CITATION = NUPHA,B158,295;%%
  
%\cite{Lindner:1985uk}
\bibitem{Lindner:1985uk} 
  M.~Lindner,
  %``Implications of Triviality for the Standard Model,''
  Z.\ Phys.\ C {\bf 31}, 295 (1986).
  %%CITATION = ZEPYA,C31,295;%%
  
%\cite{Sher:1988mj}
\bibitem{Sher:1988mj} 
  M.~Sher,
  %``Electroweak Higgs Potentials and Vacuum Stability,''
  Phys.\ Rept.\  {\bf 179}, 273 (1989).
  %%CITATION = PRPLC,179,273;%%
  
%\cite{Lindner:1988ww}
\bibitem{Lindner:1988ww} 
  M.~Lindner, M.~Sher and H.~W.~Zaglauer,
  %``Probing Vacuum Stability Bounds at the Fermilab Collider,''
  Phys.\ Lett.\ B {\bf 228}, 139 (1989).
  %%CITATION = PHLTA,B228,139;%%
  
 %\cite{Arnold:1991cv}
  \bibitem{Arnold:1991cv} 
    P.~B.~Arnold and S.~Vokos,
    %``Instability of hot electroweak theory: bounds on m(H) and M(t),''
    Phys.\ Rev.\ D {\bf 44}, 3620 (1991).
    %%CITATION = PHRVA,D44,3620;%%
    
%\cite{Altarelli:1994rb}
\bibitem{Altarelli:1994rb} 
  G.~Altarelli and G.~Isidori,
  %``Lower limit on the Higgs mass in the standard model: An Update,''
  Phys.\ Lett.\ B {\bf 337}, 141 (1994).
  %%CITATION = PHLTA,B337,141;%%  
    
%\cite{Casas:1996aq}
\bibitem{Casas:1996aq} 
  J.~A.~Casas, J.~R.~Espinosa and M.~Quiros,
  %``Standard model stability bounds for new physics within LHC reach,''
  Phys.\ Lett.\ B {\bf 382}, 374 (1996)
  [hep-ph/9603227].
  %%CITATION = HEP-PH/9603227;%%
  
%\cite{Schrempp:1996fb}
\bibitem{Schrempp:1996fb} 
  B.~Schrempp and M.~Wimmer,
  %``Top quark and Higgs boson masses: Interplay between infrared and ultraviolet physics,''
  Prog.\ Part.\ Nucl.\ Phys.\  {\bf 37}, 1 (1996)
  [hep-ph/9606386].
  %%CITATION = HEP-PH/9606386;%%
        
%\cite{Isidori:2001bm}
\bibitem{Isidori:2001bm} 
  G.~Isidori, G.~Ridolfi and A.~Strumia,
  %``On the metastability of the standard model vacuum,''
  Nucl.\ Phys.\ B {\bf 609}, 387 (2001)
  [hep-ph/0104016].
  %%CITATION = HEP-PH/0104016;%%

 %\cite{Espinosa:2007qp}
 \bibitem{Espinosa:2007qp} 
   J.~R.~Espinosa, G.~F.~Giudice and A.~Riotto,
   %``Cosmological implications of the Higgs mass measurement,''
   JCAP {\bf 0805}, 002 (2008)
   [arXiv:0710.2484 [hep-ph]].
   %%CITATION = ARXIV:0710.2484;%% 
%\cite{Clark:2009dc}
\bibitem{Clark:2009dc}
  T.~E.~Clark, B.~Liu, S.~T.~Love and T.~ter Veldhuis,
  %``The Standard Model Higgs Boson-Inflaton and Dark Matter,''
  Phys.\ Rev.\  D {\bf 80}, 075019 (2009)
  [arXiv:0906.5595 [hep-ph]].
  %%CITATION = PHRVA,D80,075019;%%

%\cite{Lerner:2009xg}
\bibitem{Lerner:2009xg}
  R.~N.~Lerner and J.~McDonald,
  %``Gauge singlet scalar as inflaton and thermal relic dark matter,''
  Phys.\ Rev.\  D {\bf 80}, 123507 (2009)
  [arXiv:0909.0520 [hep-ph]].
  %%CITATION = PHRVA,D80,123507;%%

%\cite{Gonderinger:2009jp}
\bibitem{Gonderinger:2009jp} 
  M.~Gonderinger, Y.~Li, H.~Patel and M.~J.~Ramsey-Musolf,
  %``Vacuum Stability, Perturbativity, and Scalar Singlet Dark Matter,''
  JHEP {\bf 1001}, 053 (2010)
  [arXiv:0910.3167 [hep-ph]].
  %%CITATION = ARXIV:0910.3167;%%

%\cite{EliasMiro:2011aa}
\bibitem{EliasMiro:2011aa} 
  J.~Elias-Miro, J.~R.~Espinosa, G.~F.~Giudice, G.~Isidori, A.~Riotto and A.~Strumia,
  %``Higgs mass implications on the stability of the electroweak vacuum,''
  Phys.\ Lett.\ B {\bf 709}, 222 (2012)
  arXiv:1112.3022 [hep-ph].
  %%CITATION = ARXIV:1112.3022;%%
            
%\cite{Holthausen:2011aa}
\bibitem{Holthausen:2011aa} 
  M.~Holthausen, K.~S.~Lim and M.~Lindner,
  %``Planck scale Boundary Conditions and the Higgs Mass,''
  JHEP {\bf 1202}, 037 (2012)
  [arXiv:1112.2415 [hep-ph]].
  %%CITATION = ARXIV:1112.2415;%%
              
%\cite{Xing:2011aa}
\bibitem{Xing:2011aa} 
  Z.~-z.~Xing, H.~Zhang and S.~Zhou,
  %``Impacts of the Higgs mass on vacuum stability, running fermion masses and two-body Higgs decays,''
  arXiv:1112.3112 [hep-ph].
  %%CITATION = ARXIV:1112.3112;%%

%\cite{Kadastik:2011aa}
\bibitem{Kadastik:2011aa} 
  M.~Kadastik, K.~Kannike, A.~Racioppi and M.~Raidal,
  %``Implications of the 125 GeV Higgs boson for scalar dark matter and for the CMSSM phenomenology,''
  arXiv:1112.3647 [hep-ph].
  %%CITATION = ARXIV:1112.3647;%%
                
% % Darkon Stability

\bibitem{Casas:1999cd} 
  J.~A.~Casas, V.~Di Clemente, A.~Ibarra and M.~Quiros,
  %``Massive neutrinos and the Higgs mass window,''
  Phys.\ Rev.\ D {\bf 62}, 053005 (2000)
  [hep-ph/9904295].
  %%CITATION = HEP-PH/9904295;%%
  
\bibitem{Gogoladze:2008gf} 
  I.~Gogoladze, N.~Okada and Q.~Shafi,
  %``Higgs boson mass bounds in a type II seesaw model with triplet scalars,''
  Phys.\ Rev.\ D {\bf 78}, 085005 (2008)
  [arXiv:0802.3257 [hep-ph]].
  %%CITATION = ARXIV:0802.3257;%%

\bibitem{Gogoladze:2008ak} 
  I.~Gogoladze, N.~Okada and Q.~Shafi,
  %``Higgs Boson Mass Bounds in the Standard Model with Type III and Type I Seesaw,''
  Phys.\ Lett.\ B {\bf 668}, 121 (2008)
  [arXiv:0805.2129 [hep-ph]].
  %%CITATION = ARXIV:0805.2129;%%
  
%\cite{Hambye:1996wb}
\bibitem{Hambye:1996wb} 
  T.~Hambye and K.~Riesselmann,
  %``Matching conditions and Higgs mass upper bounds revisited,''
  Phys.\ Rev.\ D {\bf 55}, 7255 (1997)
  [hep-ph/9610272].
  %%CITATION = HEP-PH/9610272;%%  
  
%\cite{Cheng:1973nv}
\bibitem{Cheng:1973nv} 
  T.~P.~Cheng, E.~Eichten and L.~-F.~Li,
  %``Higgs Phenomena in Asymptotically Free Gauge Theories,''
  Phys.\ Rev.\ D {\bf 9}, 2259 (1974).
  %%CITATION = PHRVA,D9,2259;%%

%\cite{Machacek:1983tz}
\bibitem{Machacek:1983tz}
  M.~E.~Machacek and M.~T.~Vaughn,
  %``Two Loop Renormalization Group Equations In A General Quantum Field Theory.
  %1. Wave Function Renormalization,''
  Nucl.\ Phys.\  B {\bf 222}, 83 (1983).
  %%CITATION = NUPHA,B222,83;%%

%\cite{Machacek:1983fi}
\bibitem{Machacek:1983fi}
  M.~E.~Machacek and M.~T.~Vaughn,
  %``Two Loop Renormalization Group Equations In A General Quantum Field Theory.
  %2. Yukawa Couplings,''
  Nucl.\ Phys.\  B {\bf 236}, 221 (1984).
  %%CITATION = NUPHA,B236,221;%%

%\cite{Machacek:1984zw}
\bibitem{Machacek:1984zw}
  M.~E.~Machacek and M.~T.~Vaughn,
  %``Two Loop Renormalization Group Equations In A General Quantum Field Theory.
  %3. Scalar Quartic Couplings,''
  Nucl.\ Phys.\  B {\bf 249}, 70 (1985).
  %%CITATION = NUPHA,B249,70;%%  
      
\bibitem{Arason:1991ic} 
  H.~Arason, D.~J.~Castano, B.~Keszthelyi, S.~Mikaelian, E.~J.~Piard, P.~Ramond and B.~D.~Wright,
  %``Renormalization group study of the standard model and its extensions. 1. The Standard model,''
  Phys.\ Rev.\ D {\bf 46}, 3945 (1992).
  %%CITATION = PHRVA,D46,3945;%%      
  

% % neutrino beta function
%\cite{Chakrabortty:2008zh}
\bibitem{Chakrabortty:2008zh} 
  J.~Chakrabortty, A.~Dighe, S.~Goswami and S.~Ray,
  %``Renormalization group evolution of neutrino masses and mixing in the Type-III seesaw mechanism,''
  Nucl.\ Phys.\ B {\bf 820}, 116 (2009)
  [arXiv:0812.2776 [hep-ph]].
  %%CITATION = ARXIV:0812.2776;%%  


  
\end{thebibliography}
\end{document}